\documentclass[aps, prl, superscriptaddress, twocolumn,showpacs]{revtex4}
\usepackage{graphicx}
\usepackage{amsmath}
\usepackage{amssymb}
\usepackage{bm}

\begin{document}
\setlength{\unitlength}{1mm}


\title{Dynamics of drag and force distributions for projectile impact in a granular medium}

\author{Massimo Pica Ciamarra}%
\email[]{picaciamarra@na.infn.it}

\affiliation{Center for Nonlinear Dynamics and Department of
Physics, University of Texas
at Austin, Austin, Texas 78712}%
\affiliation{Dipartimento di Scienze Fisiche, Universit\'a di
Napoli `Federico II' and INFM, Unit\'a di Napoli, 80126 Napoli,
Italia.}
\author{Antonio H. Lara}
\affiliation{Center for Nonlinear Dynamics and Department of
Physics, University of Texas
at Austin, Austin, Texas 78712}%
\author{Andrew T. Lee}
\affiliation{Center for Nonlinear Dynamics and Department of
Physics, University of Texas
at Austin, Austin, Texas 78712}%
\author{Daniel I. Goldman}
\affiliation{Center for Nonlinear Dynamics and Department of
Physics, University of Texas
at Austin, Austin, Texas 78712}%
\author{Inna Vishik}
\affiliation{Center for Nonlinear Dynamics and Department of
Physics, University of Texas
at Austin, Austin, Texas 78712}%
\author{Harry L. Swinney}
\email[]{swinney@chaos.utexas.edu} \affiliation{Center for
Nonlinear Dynamics and Department of Physics, University of Texas
at Austin, Austin, Texas 78712}%
\date{\today}

\begin{abstract}

% 599 characters with spaces; 600 allowed:
Our experiments and molecular dynamics simulations on a projectile
penetrating a two-dimensional granular medium reveal that the mean
deceleration of the projectile is constant and proportional to the
impact velocity. Thus, the time taken for a projectile to
decelerate to a stop is independent of its impact velocity. The
simulations show that the probability distribution function of
forces on grains is time-independent during a projectile's
deceleration in the medium. At all times the force distribution
function decreases exponentially for large forces.

\end{abstract}
%\pacs{45.70.Ht, 4.70.Cc, 83.30.Fg, 89.75.Da}

\maketitle

Craters on the earth and moon are similar to craters formed in
laboratory experiments at much lower energies by using projectiles
and explosives \cite{Roddy,Mi83,Mel}.  In laboratory experiments
at large impact energies, narrow jets have been observed to rise
even higher than the initial height of the projectile
\cite{Th01,Lo02}. Recent experiments have determined how the
shape, depth, and width of craters formed in granular media depend
on the energy of the impact projectile \cite{Du03,DeB03}, but
there is little known about the dynamics of a projectile during
crater formation.

We have studied the time evolution of projectile motion. Our
experiments and molecular dynamics simulations on a
two-dimensional granular medium yield the time dependence of the
drag force on projectiles. Simulations for the same conditions are
in accord with the experiment and also yield the time evolution of
the forces on all of the particles; hence, we can study the time
dependence of the force probability distribution function at
different stages of the projectile motion.

\begin{figure}
  \begin{tabular}{cccc}
    & Impact & Penetration & Collapse \\
    \rotatebox{90}{\hspace{0.6cm} Experiment} &
    \includegraphics[scale = 0.21]{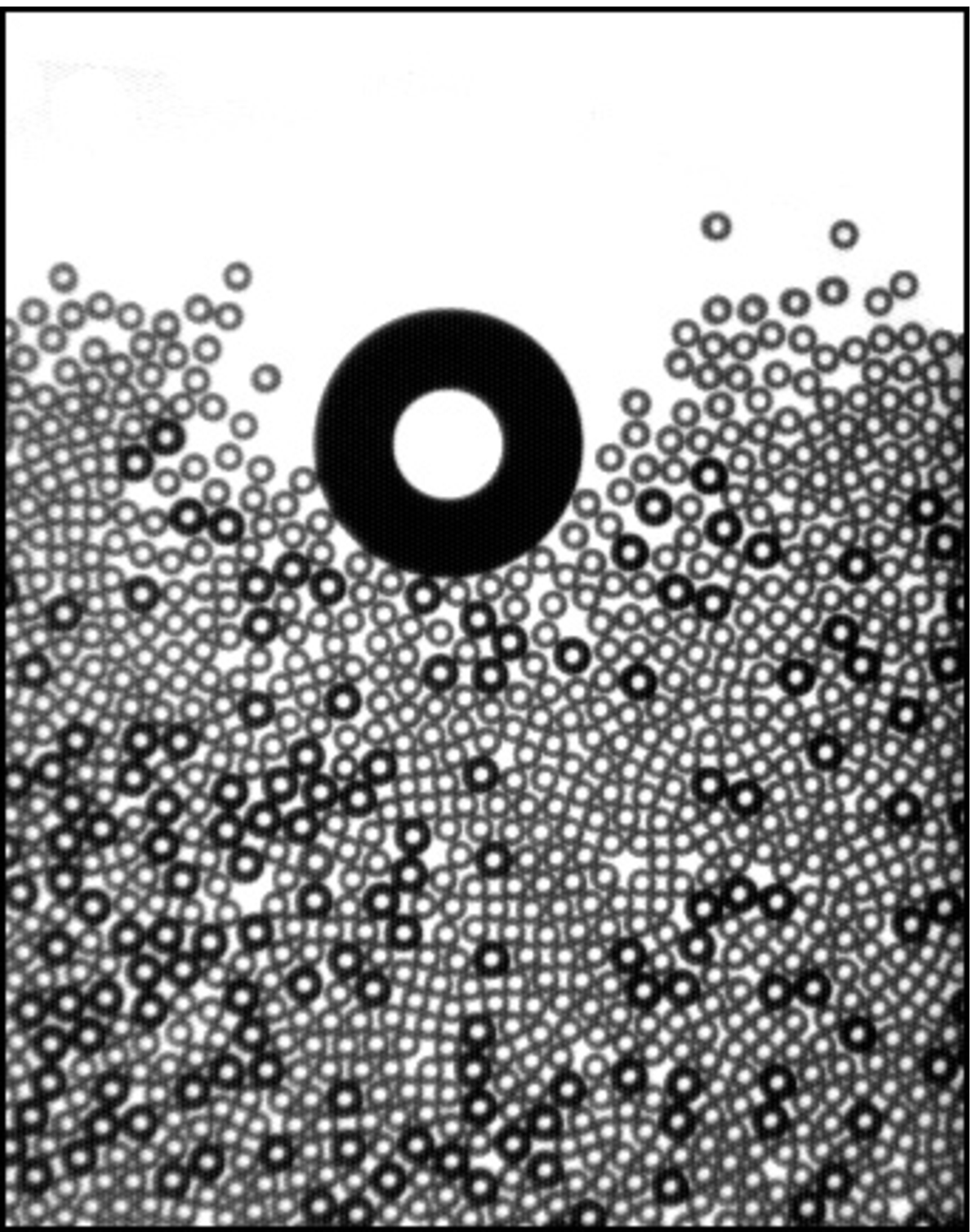}
    &
    \includegraphics[scale = 0.21]{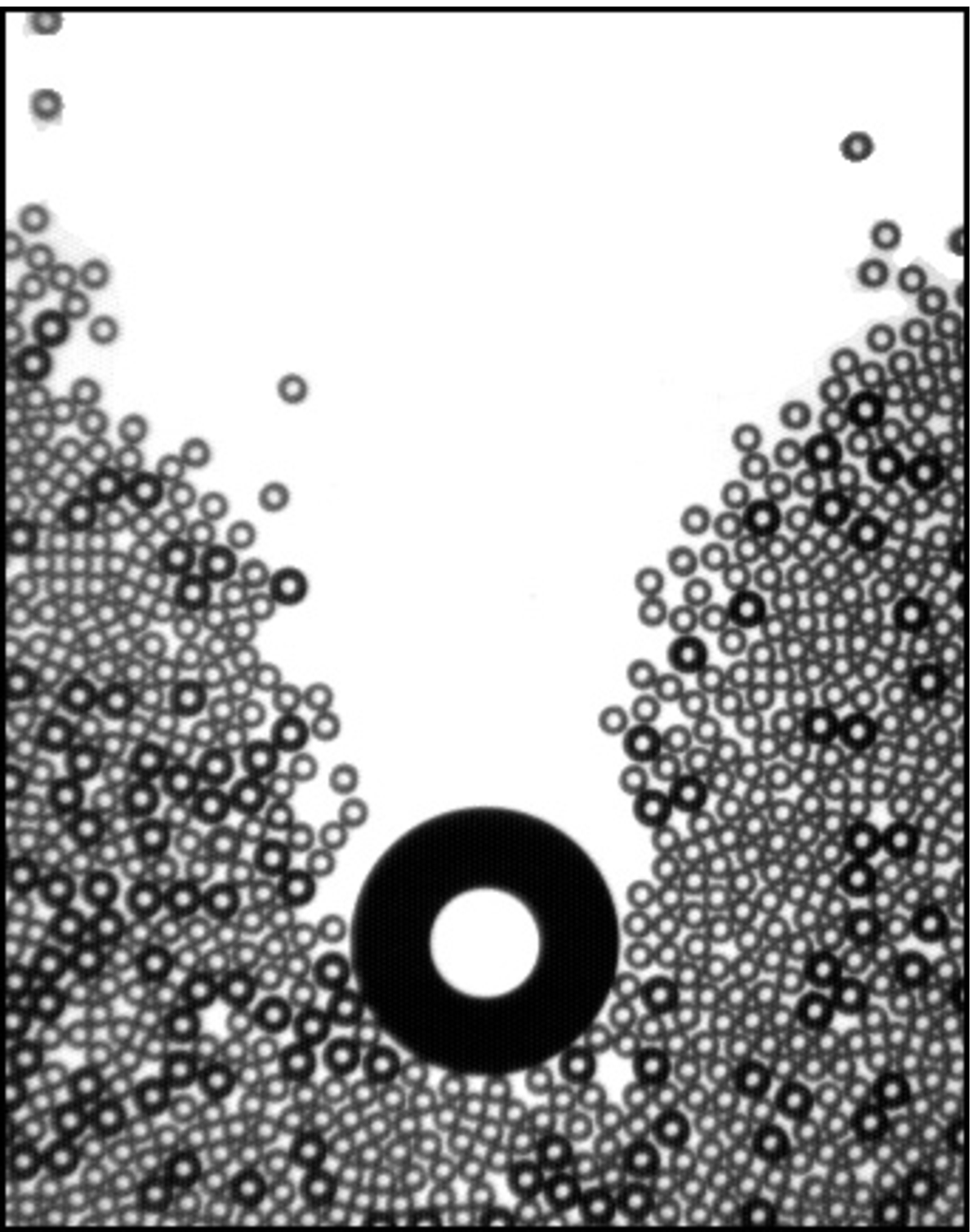}
    &
    \includegraphics[scale = 0.21]{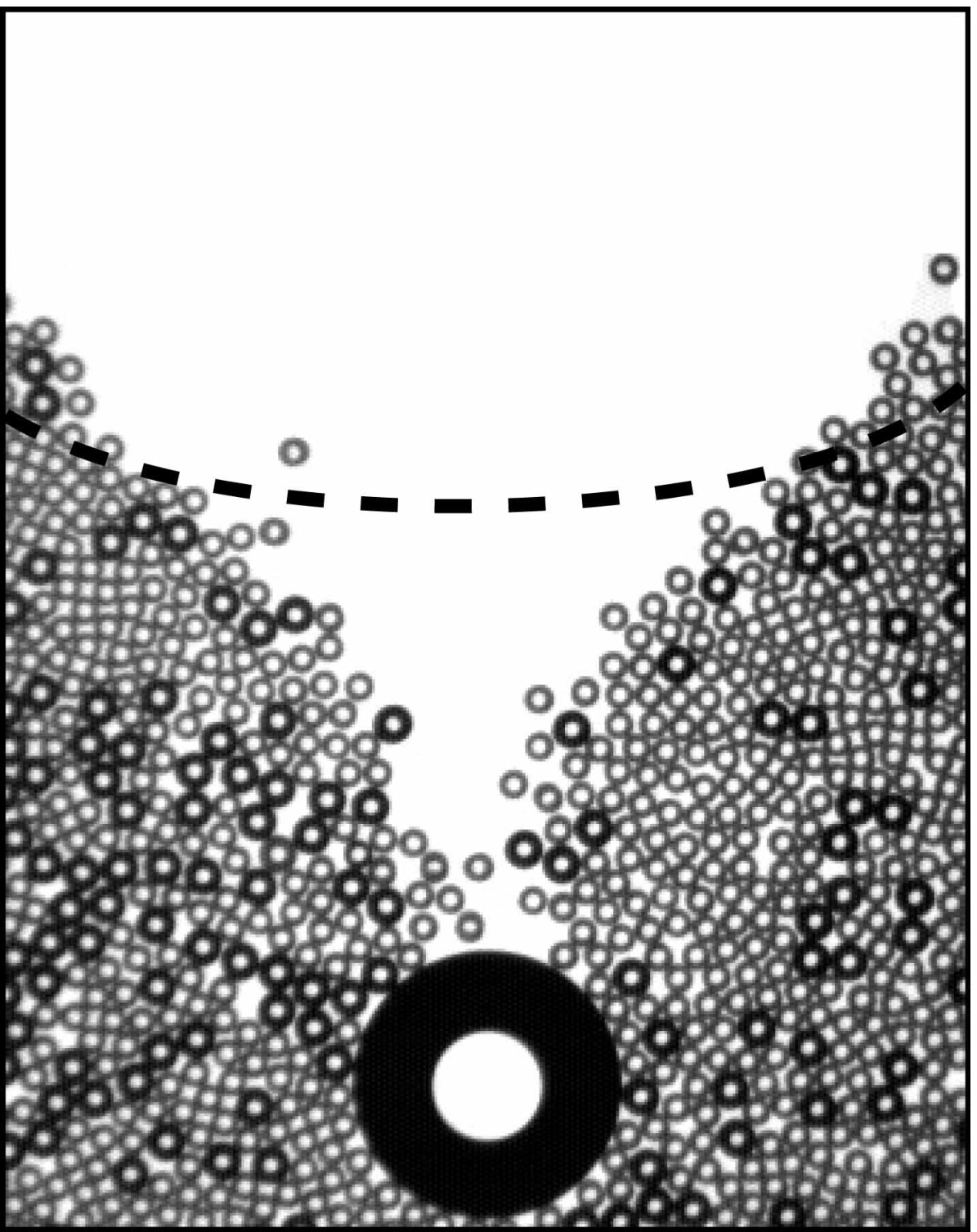}
    \vspace{-.12in}
    \\
    \rotatebox{90}{\hspace{-2.7cm} Simulation}
    &
    \includegraphics[scale = 0.33, angle = 270]{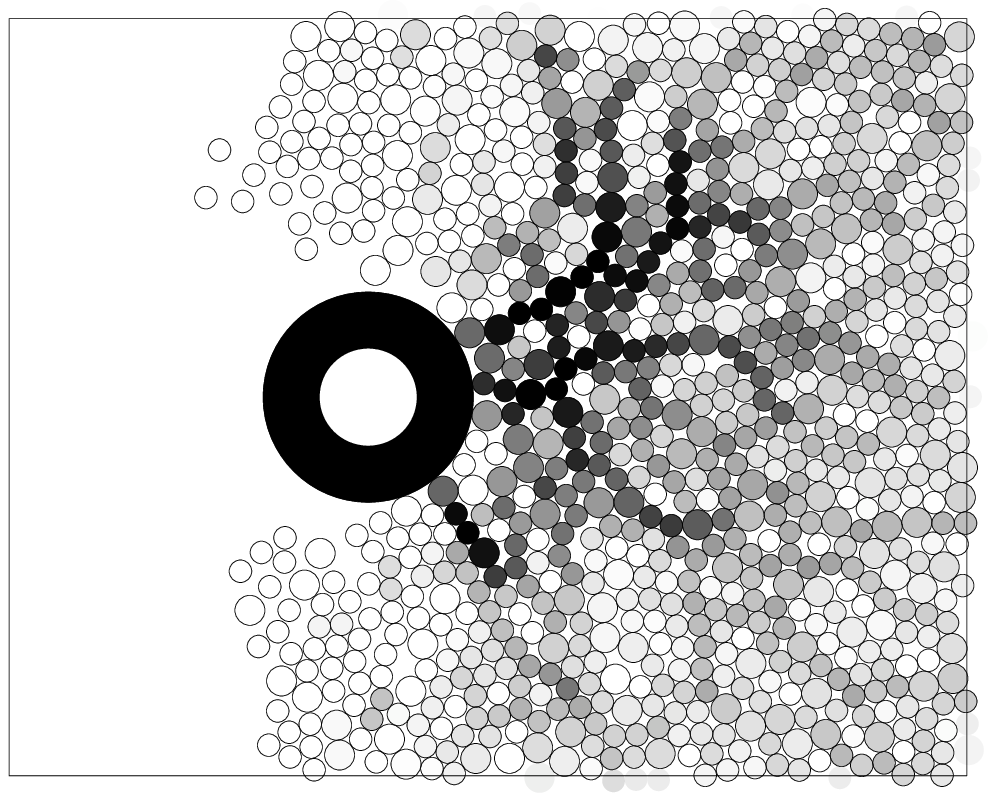}
    &
    \includegraphics[scale = 0.33, angle = 270]{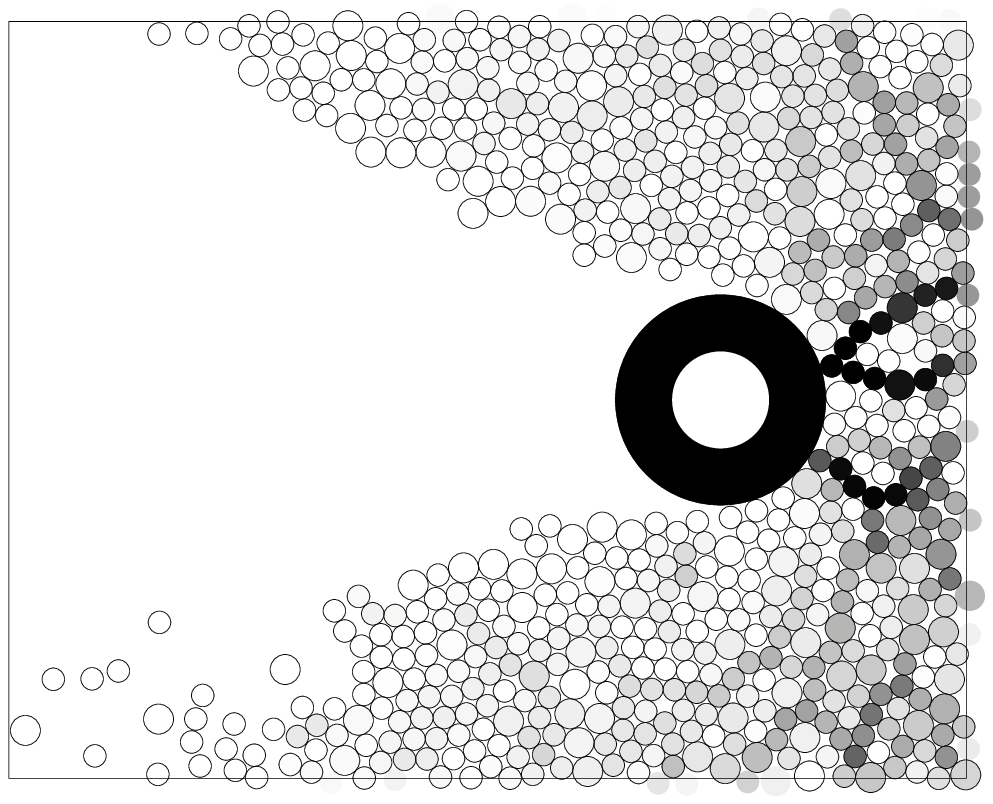}
    &
    \hspace{-.04in}\includegraphics[scale = 0.33, angle = 270]{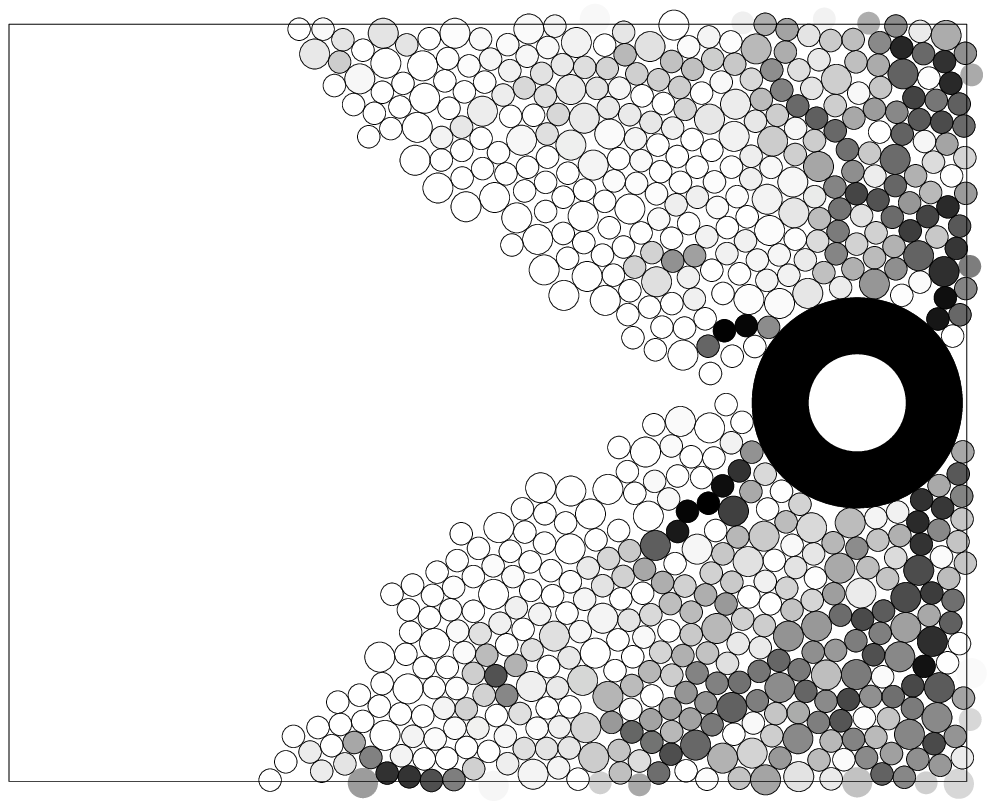}
    \\
     & $0.016$ s & $0.08$ s & $0.15$ s
  \end{tabular}
 \caption{
Snapshots of a projectile in the three distinct regimes of its
motion in a bidisperse mixture of particles (cylinders).
Experiment: The larger cylinders of the bidisperse mixture are
colored black for visualization and are 40\% larger in diameter
than the grey cylinders.  The dashed line shows the location of
the surface after the collapse is complete.  Simulation: the
shading of each particle is proportional to the sum of the
magnitudes of all the normal forces acting on that particle; this
renders visible the instantaneous force chains. The projectile is
9.8 times as large in diameter and 657 times as massive as the
smallest particles.}

  \label{fig1}
\end{figure}

Our observations and simulations reveal three distinct regimes of
the motion, as illustrated in Fig.~\ref{fig1}: {\it impact}, where
the projectile first hits the granular medium; {\it penetration},
where a transient crater forms and grains in front of the
projectile are fluidized; {\it collapse}, where the projectile has
almost stopped and the deep transient crater collapses, forming a
static crater that remains visible on the surface.

\textit{Methods} --- In the experiment, a projectile of diameter
$D=4.46$ cm and mass $32.2$ g was dropped into a bed of small
particles (cylinders) contained between two glass plates with a
separation $1.1$ times the length of the cylinders.  The initial
projectile heights $h$ ($h < 80$ cm) correspond to impact
velocities up to $400$ cm/s.  To reduce crystallization, two sizes
of small particles were used: 12600 particles (84\% of the total
number) had diameter $d_1 = 0.456$ cm (mass $m_1 = 0.049$ g) and
2400 particles had diameter $d_2=0.635$ cm (mass $m_2 = 0.097$ g).
To obtain a uniform granular bed with a reproducible area fraction
before each drop of the projectile, the bed was fluidized with air
flow that was slowly reduced to zero, yielding the same bed height
(65$d_1$) and area fraction ($81\pm2$\%) for each projectile drop.
The bed width was 225$d_1$. The position of the projectile,
$y(t)$, defined as the distance between the bottom of the
projectile and the initial height of the bed, was determined with
a high speed camera and a center of mass particle tracking
algorithm \cite{Crocker}.

We modelled the system with a soft-core molecular dynamics (MD)
simulation that used $15,000$ disks that had the same sizes and
area fraction as the experiment.  Any two disks (one of which can
be the projectile) exert the following normal and tangential
forces on one another:
\begin{eqnarray}
\vec F_n = -\left[ k \delta + m_r \gamma_n |\vec v_n| \theta(\vec v_n) \right] \hat n \\
\vec F_s = \min\left[ m_r \gamma_s |\vec v_s| ,\mu |\vec
F_n|\right] \hat s, \label{eq-simulation}
\end{eqnarray}
where $\delta$ is the length of overlap \cite{Bu98,Bri96}, and
$\vec v_n$ and $\vec v_s$ are the normal and tangential components
of the surface velocity ($\hat n$ and $\hat s$ are unit vectors
parallel to $\vec v_n$ and $\vec v_s$). The four parameters of the
model were found empirically for one impact velocity and the same
parameters were used for all other simulations: $k = 3.2\times
10^3$ kg s$^{-2}$ \cite{ft1,Sib01,Rap02,La03} is proportional to
Young's modulus, $\gamma_n = 10^4$ s$^{-1}$ and $\gamma_s =
8\times 10^3$ s$^{-1}$ are viscoelastic constants, and $\mu =
0.28$ is the static friction coefficient.  $m_r$ is the reduced
mass ($m_r^{-1} = m_A^{-1} + m_B^{-1}$ for two particles A and B).
The Heaviside function $\theta$ in $\vec F_n$ models an
elastic-plastic interaction (e.g., see Fig. 8 of \cite{Ro97}); the
use of the Heaviside function distinguishes our force model from
previous soft-core MD simulations \cite{Bu98, La03}. Simulations
with a more realistic form for $\vec F_s$ \cite{Cu79} yielded
results not significantly different from our simple form, which is
computationally more efficient.  A comparison of the simulation
output using time steps shorter than $1$ $\mu$s did not yield
different results; a $1$ $\mu$s time step was used in the results
presented here.

\begin{figure}[t]
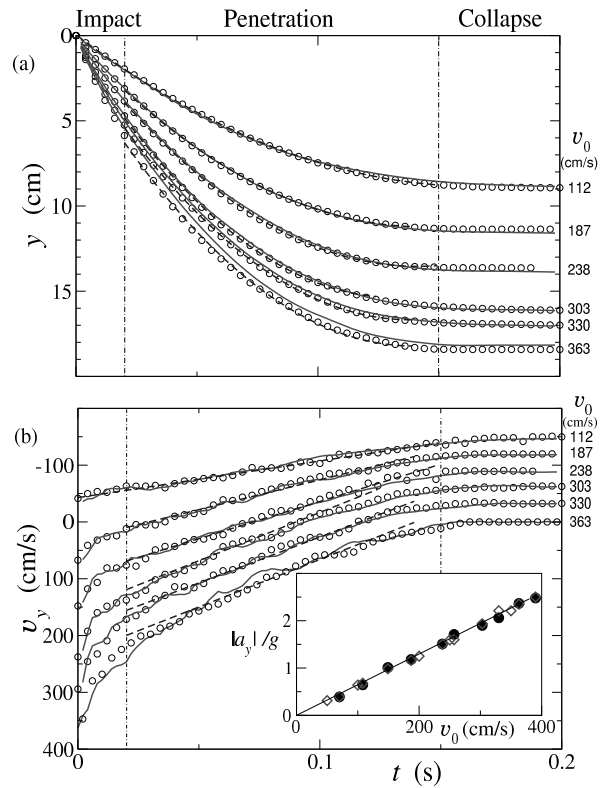

\begin{tabular}{c}
\includegraphics*[scale = 0.30]{pos_paper_gray.eps}
%\vspace{-.5in}
\vspace{-.03in}
\\
\hspace{-.08in}
\includegraphics*[scale = 0.30]{vel_paper_gray_no_label.eps}
\end{tabular}
\caption{(a) Position $y(t)$ and (b) velocity $v_y(t)$ of the
projectile as a function of time for different impact velocities,
from both experiment ($\circ$) and simulation (solid lines). The
two vertical dot-dashed lines give approximate boundaries between
three regions: impact, where the projectile rapidly decelerates
(cf. Fig.~\ref{fig-spectrum}); penetration, where the mean
acceleration is constant, as illustrated by a dashed line fit in
(a) of a parabola to the results from experiment and simulation
for each $v_0$; and collapse, where the projectile has almost
stopped and the particles above it are collapsing to fill the
transient crater left by the penetration. The ordinate for (b) for
each successive impact velocity
 $v_0 < 363$ cm/s is shifted by $30$ cm/s for clarity. Inset: normalized
acceleration of the projectile versus impact velocity from
experiment ($\bullet$) and simulation ($\diamondsuit$).}
\label{fig2}
\end{figure}

\begin{figure}
%\vspace{-.35in}
\includegraphics*[scale=0.35]{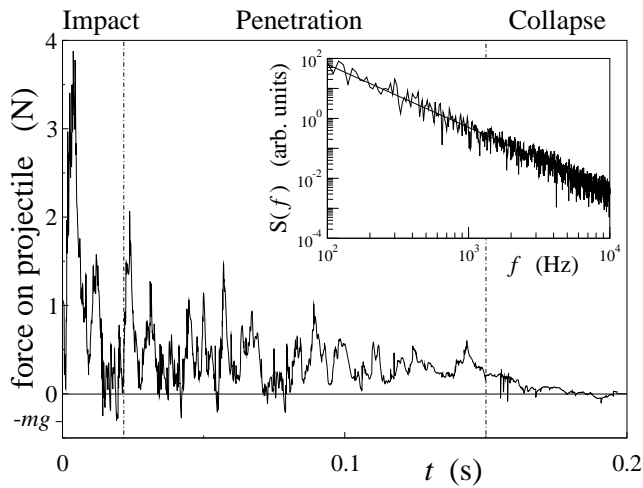}
\vspace{-.0in} \caption{The time series of the force on the
projectile obtained from the simulation.  The three regimes of
motion are separated by dot-dashed lines.  Inset: The power
spectrum of the projectile acceleration during the penetration
regime (0.02-0.15 s) for a projectile with initial velocity
$v_0=238$ cm/s is described by $f^{-\alpha}$ with $\alpha=2.1\pm
0.2$. } \label{fig-spectrum}
\end{figure}

\textit{Results} --- The simulation results agree remarkably well
with the laboratory observations, as Fig.~\ref{fig2} illustrates.
Both experiment and simulation reveal that the time taken for a
projectile to slow to a stop in the granular medium is {\it
independent} of its velocity at impact. The large deceleration of
the projectile at impact (see Fig.~\ref{fig-spectrum}) is similar
to that of a projectile incident on a liquid. However, in contrast
to the behavior of a projectile in a fluid \cite{gl96}, in the
granular medium there is a long penetration region in which the
projectile's average acceleration is constant: $y(t)$ is described
by a parabola (Fig.~\ref{fig2}(a)), so $v_y(t)$ decreases linearly
in time (Fig.~\ref{fig2}(b)). Further, the acceleration is
proportional to the impact velocity, as the inset in
Fig.~\ref{fig2}(b) illustrates: $a_y = \alpha v_0 g$, where the
slope of the line gives $\alpha=0.0064\pm0.0001$ s/cm. Thus, the
projectile slows almost to a stop in a time $t=1/\alpha g \simeq
0.15$ s, independent of $v_0$. The projectile does not immediately
come to a complete stop; rather it then moves very slowly downward
over the next few seconds as the particles in the bed make small
rearrangements in response to the collapse of the transient
crater.

The drag force on the projectile, while constant on the average,
exhibits large fluctuations, which have a $f^{-2}$ spectrum
(Fig.~\ref{fig-spectrum}).

\begin{figure}[t]
\includegraphics*[scale=0.35]{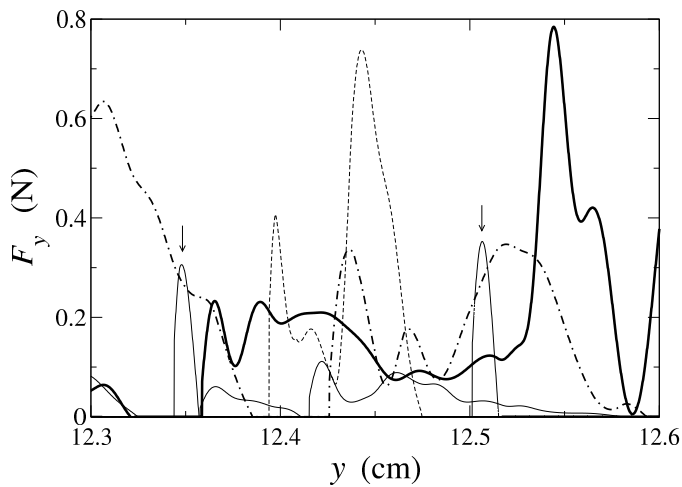}
\caption{Vertical component of the force computed for {\it every}
particle in contact with the projectile during part of the
penetration regime ($0.100 < t < 0.108$ s in Figs.~\ref{fig2}
and~\ref{fig-spectrum}). Each force grows, reaches a maximum
(representing the inclusion of a particle in a particular force
chain), and then decreases.  Each type of line represents a
particular particle; thus, the two arrows correspond to the {\it
same} particle that appeared first at 12.344 cm and then
reappeared at 12.501 cm. The projectile impact velocity was $v_0 =
238$ cm/s. The average of the total force on the projectile during
this interval was $0.57$ N. } \label{fig-forces}
\end{figure}

The simulation determines all of the forces on each particle at
every instance of time.  Every force exerted by a particle on the
projectile during a short portion of its travel is shown in
Fig.~\ref{fig-forces}.  At each point in the projectile's
trajectory only a few particles exert a significant force on the
projectile. Each peak in the magnitude of the force between an
individual particle and the projectile in Fig.~\ref{fig-forces}
corresponds to a maximum force felt by the first particle in a
force chain \cite{Al00} that extends downward.  Each force chain
consists of a string of particles in contact.  The sum of the
magnitudes of forces felt by each particle in this chain is much
greater than the average for the particles in the bed, as can be
seen in Fig.~\ref{fig1} (simulation), where dark chains of
particles extend downward from the projectile into the particle
bed.

\begin{figure}[t]
\includegraphics*[scale = 0.35]{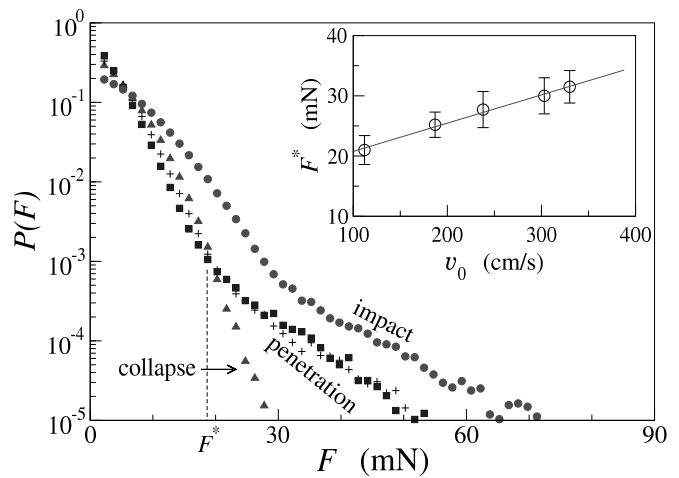}
\caption{Probability distribution of normal contact forces between
grains for a projectile with $v_0 = 112$ cm/s at the following
times: during impact ($t_1 = 0.02 $s, $\bullet$), early in the
penetration regime ($t_2 = 0.05$ s, $\blacksquare$), late in the
penetration regime ($t_3 = 0.12 $ s, $+$), and during collapse
($t_4 = 0.20$ s, $\blacktriangle$). The distribution decays
exponentially for $F > F^*$. The dependence of $F^*$ on the impact
velocity is shown in the inset; the slope is $0.047 \pm 0.004$
mN$\cdot$s/cm. Each curve was obtained by averaging over $50$
runs. } \label{fig-prob}
\end{figure}

Results for the probability distribution $P(F,t)$ of all normal
forces between particles located in front of the projectile in a
semicircular region of radius $1.5D$ centered at the bottom-most
point of the projectile are shown in Fig.~\ref{fig-prob}.  The
distribution $P(F,t)$ changes with time during impact but is time
invariant during penetration: Fig.~\ref{fig-prob} shows the same
distribution at times $t_2$ and $t_3$, which are respectively
early and late in the penetration regime. The presence of an
inflection point $F^*$ in $P(F,t)$ marks the beginning of
exponential decay for large $F$.  The cross-over to an exponential
distribution at $F^*$ increases linearly with $v_0$, as shown in
the inset of Fig.~\ref{fig-prob}. After the projectile has almost
stopped, the distribution is similar to that found in previous
studies of equilibrium \cite{ohe20} and near equilibrium
\cite{Ho99} force distributions .

{\it Discussion} ---  Our experiments and simulations show that
the mean drag force on a projectile dropped into a granular medium
is constant during most of the projectile's trajectory, and this
drag force is proportional to the projectile's impact velocity. In
our experiments inertia plays a major role. Interestingly,
previous experiments with low constant velocities and negligible
inertial effects also yielded a constant drag force in a granular
medium\cite{Alb99}.

Since the deceleration of the projectile is proportional to the
impact velocity (see inset Fig.~\ref{fig2}(b)), the projectile
penetration depth is also proportional to the impact velocity.
While our results are for a two-dimensional system, the linear
dependence of the penetration depth on impact velocity has
recently also been observed for projectile impact in a
three-dimensional granular medium \cite{DeB03b}.

The drag force on our projectile fluctuates strongly, as found
also for cylinders dragged at small constant velocities in
experiments ($v \simeq 0.1$ cm/s) \cite{Alb99} and simulations ($v
\simeq 2$ cm/s)\cite{No20,ft2,Bu98}. The power spectrum of the
force fluctuations has a $f^{-2}$ dependence, as observed in
measurements of fluctuations of the stress on a slowly sheared
two-dimensional granular medium~\cite{Beh96} and in measurements
of the torque on a torsional pendulum in contact with a
vibrofluidized granular bed~\cite{danna01}. The $f^{-2}$
dependence is explained by assuming random jumps in the drag
force~\cite{Beh96}. In our experiment these jumps originate from
the variation of the forces exerted by the grains in contact with
the projectile (Fig.~\ref{fig-forces}).

Finally, our simulations have yielded the normal contact forces
for all particles in the bed.  The distribution function for the
forces on the particles in front of the projectile rapidly evolves
immediately after the projectile makes contact with the bed, and
then the distribution becomes stationary as the projectile
penetrates the bed. This stationary distribution decays
exponentially beyond an inflection point at $F^*$ whose value is
linearly proportional to the impact velocity. This is the first
determination of the force distribution for a granular medium for
an accelerating particle. During impact, our force distribution is
different from that measured for static beds \cite{ohe20}, where
the force distribution decayed exponentially at all times, as
predicted by the $q$-model~\cite{Co95}.

\begin{acknowledgments}
We thank John de Bruyn and W. D. McCormick for their helpful
comments and suggestions. This work was supported by the
Engineering Research Program of the Office of Basic Energy
Sciences of the U. S. Department of Energy (Grant No.
DE-FG03-93ER14312), the Texas Advanced Research Program, and the
Office of Naval Research Quantum Optics Initiative.  M.P.C.
gratefully acknowledges support of the Italian-Fulbright
commission.
\end{acknowledgments}
%\bibliography{article}

\begin{thebibliography}{26}
\bibitem[{ft1()}]{ft1}
\bibinfo{note}{As in most other MD simulations of granular media with a soft
  core potential (e.g. \cite{Sib01,Rap02}), we use a value of $k$ that is far
  smaller than the physical value ($10^6$ kg s$^{-2}$ for nylon) because a
  higher value is computationally too expensive; the integration time must have
  the form $\delta t \varpropto k^{-1/2}$ for collisions to be modelled
  effectively \cite{La03}. The success of past \cite{Sib01,Rap02} and present
  MD simulations with a small value of $k$ indicates that the model, despite
  this flaw, captures much of the dissipative dynamics.}
\bibitem[{ft2()}]{ft2}
\bibinfo{note}{For a projectile penetrating a dilute granular medium at high
  velocity ($v > 10^2$ cm/s)\cite{Bu98}, much smaller fluctuations than we
  observe have been found.}
\end{thebibliography}

\end{document}